\documentstyle[aps,psfig]{revtex}

\newcommand{\be}{\begin{equation}}
\newcommand{\ee}{\end{equation}}

\newcommand{\ber}{\begin{eqnarray}}
\newcommand{\eer}{\end{eqnarray}}

\begin{document}
\draft
\title {\bf Rotating Quark Star in Chiral Colour Dielectric Model}
\vskip 0.2in
\author{Abhijit Bhattacharyya$^a$\footnote
{E-Mail :bhattacharyyaabhijit\_10@yahoo.co.uk} and Sanjay K. Ghosh$^b$\footnote
{E-Mail : sanjay@bosemain.boseinst.ac.in} }
\address{$^a$ Department of Physics, Scottish Church College, 1 \& 3, Urquhart 
Square, Kolkata - 700 006, INDIA}
\address{$^b$ Department of Physics, Bose Institute, 93/1, A.P.C. Road,
Kolkata - 700 009, INDIA}
\maketitle
\vskip 0.4in
\begin{abstract}
The properties of rotating quark star is studied using the
equation of state obtained from Chiral Colour Dielectric model. The results are
compared with the MIT bag model results. The frequencies
in the corotating innermost circular orbits for different central densities are
evaluated and compared with the observational results.
\end{abstract}
\vskip 0.3in
\section{Introduction}
The quark structure of hadrons suggests the possibility of a phase
transition from nuclear to quark matter at high density. Since the inception
of the concept of phase transition to strange quark matter (SQM) at 
high density and the 
possibility of the existence of exotic compact objects \cite{witten}, several
calculations have been done to investigate the existence of quark stars 
\cite{alcock} or quark core in neutron stars using different models 
\cite{collins}. The phase transition to strange quark matter has been shown
\cite{sannpa} to result in the production of large amount of energy along 
with the neutrinos. But an unambiguous identification of such processes, 
for example, in the form of gamma-ray bursters has not been possible till date. 
On the other hand, Glendenning {\it et al.} \cite{glend1} have argued that 
anomalous behaviour of braking index may be a signature of phase transition 
to quark matter. Subsequently, studies of braking index using different 
models \cite{steri1,abhi1} has been reported. Twin star solutions has also
been obtained for both static as well as rotating stars \cite{abhi1}
-\cite{abhi3}.  

At present, despite various studies, 
there is no consensus regarding the existence of quark stars in the universe. 
Most of the calculations on the static properties of quark star do not 
yield any significant observable to distinguish them
from neutron stars. 
\par
The study of exotic compact objects, like quark stars, has once again
become important in the context of compact X-ray and \( \gamma \)-ray 
sources \cite{bombaci}. Several studies on rotating compact objects 
have reported a substantial difference in the properties of rotating strange 
quark and neutron star, attributed mainly to their different equation of 
state \cite{gour,zdunik1}. In particular, qualitative differences 
in the properties of the innermost stable circular orbits (ISCO) of
strange quark stars and neutron stars have been found \cite{steri2,zdunik1}.
Moreover neutron star calculations yield lower values for ISCO frequencies
compared to the quark stars \cite{limou}. 
It has been shown that a comparison of the theoretical value obtained for 
the ISCO with the kHz quasi periodic oscillations (QPO) found in low mass 
X-ray binaries (LMXB) can be used to constrain the SQM models. 
\par
In the present paper we have studied the rotating strange quark star in
the general relativistic framework using the nonlinear Chiral extension
of Colour Dielectric Model. 
The  Chiral Colour Dielectric model (CCDM) has been used earlier to study 
baryon spectroscopy \cite{sarira} as well as the static 
properties of nucleons in nuclear medium \cite{phatak}. These calculations 
have shown that the model is able to explain the static properties of 
light baryons very well. Furthermore, when applied to the quark matter 
calculation, the model yields an 
equation of state which is quite similar to the one obtained from lattice 
calculations for zero baryon chemical potential \cite{san1}. It has also 
been used for the calculation of the properties of 
dibaryons \cite{san2}, static hybrid stars (neutron stars with quark core) 
\cite{san3} and static strange quark stars \cite{san4}. 
The  CCDM differs from the bag model in several aspects. First of all, 
in the CCDM, the confinement of quarks and gluons is achieved 
dynamically through the colour-dielectric field. In the bag model this 
is done by hand. Also the quark masses used in the CCDM are different 
from those used in the bag model. In the bag model, \( u \) and \( d \) 
masses are taken to be zero. The CCDM requires that these masses are 
nonzero. It has been found \cite{sarira}
that to fit baryon masses, the required \( u \) and \( d \) masses are 
\( \sim 100MeV \). Thus, the values of quark masses in CCDM
are closer to the constituent quark masses. Motivated by the earlier 
successes of the CCDM we have used this model for the study of
rotating quark star. 
\par
The paper is organized as follows. In the section 2 a brief description of 
the CCDM is presented. 
Calculations for the rotating star is described in
section 3 followed by a summary in the last section. 
\section{Chiral Colour Dielectric Model}
The Lagrangian density of CCDM
is given by \cite{san1}
\begin{eqnarray}
{\cal{L}}(x)&= \bar\psi(x)\big \{ i\gamma^{\mu}\partial_{\mu}-
(m_{0}+m/\chi(x) U_{5}) + (1/2) g
\gamma_{\mu}\lambda_{a}A^{a}_{\mu}(x)\big \}\psi \nonumber \\
&+f^{2}_{\pi}/4 Tr ( \partial_{\mu}U
\partial^{\mu}U^{\dagger} ) -
1/2m^{2}_{\phi} \phi^{2}(x) -(1/4)
\chi^{4}(x)(F^{a}_{\mu\nu}(x))^{2} \nonumber \\ &+ (1/2)
\sigma^{2}_{v}(\partial_{\mu}\chi(x))^{2}- U(\chi)
\end{eqnarray}

where \( U = e^{i\lambda_{a}\phi^{a}/f_\pi} \) and \( U_{5} =
e^{i\lambda_{a}\phi^{a}\gamma_{5}/f_{\pi}}\), \( \psi(x) \),
\( A_{\mu}(x) \), and \( \chi(x) \) and \( \phi(x) \) are quark, 
gluon, scalar ( colour dielectric )and meson fields respectively.
The quark and meson masses are denoted by \( m \) and \( m_{\phi}\) 
respectively, \( f_{\pi} \) is the pion
decay constant, \( F_{\mu\nu}(x) \) is the usual colour
electromagnetic field tensor, g is the colour coupling constant
and \( \lambda_{a} \) are the Gell-Mann matrices.  The flavour
symmetry breaking is incorporated in the Lagrangian through the
quark mass term \( (m_{0}+m/\chi U_{5}) \), where \( m_0= 0 \) 
for u and
d quarks. So  masses of u, d and s quarks are \( m \), \( m \) and
\( m_{0}+ m \) respectively. So for a system with
broken flavour symmetry, strange quark mass will be different from 
u and d quark masses. 
The meson matrix then consists of a singlet \( \eta \), triplet of
\( \pi \) and quadruplet of \( K \). 
With \( m_0 = 0 \), one can recover a symmetric
three flavour quark matter system.
The corresponding meson matrix \( \Phi \)
then becomes a eight component field. 
\par
The self interaction
\( U(\chi) \) of the scalar field is assumed to be of the form
\begin{eqnarray}
U(\chi)~=~\alpha B
\chi^2(x)[
1-2(1-2/\alpha)\chi(x)+(1-3/\alpha)\chi^2(x)]
\end{eqnarray}
so that \( U(\chi) \) has an absolute minimum at \( \chi = 0 \) and a 
secondary
minimum at \( \chi = 1 \). The interaction of the scalar field with quark
and gluon fields is such that quarks and gluons can not exist in the
region where \( \chi= 0 \).
In  the  limit  of  vanishing  meson  mass,  the
Lagrangian of eqn.(1) is invariant under chiral transformations of
quark and meson fields.
\par
In general there are approaches which can be followed in the 
studies of hadronic systems using chiral models.
One is the perturbative methodology of cloudy bag model 
\cite{theberg}. The another one 
is hedgehog approach \cite{chodos}, which is nonperturbative but 
is not applicable for infinite matter. A different ansatz
has been proposed in ref. \cite{san1} in an 
attempt to go beyond the perturbative approach of cloudy bag model, 
One can assume that because of nonvanishing quark and antiquark densities, 
the square of the expectation value of meson fields develop a nonzero
value i.e. $<\phi^{2}>\not = 0$. On the other hand it is assumed 
that the expectation value of the meson field vanishes in the medium
. For an infinite system of quarks one can take 
$<\phi^{2}>$ to be independent of space and time. The meson
excitations are then defined in terms of the fluctuations about
$<\phi^{2}>$, so that $\phi^{2}~=~<\phi^{2}>~+ ~{\phi^{'}}^{2}$.
Defining $F_{\phi}~= <\phi^{2}>/f_{\pi}^{2}$, the CCDM Lagrangian
can be rewritten in terms of $F_{\phi}$s and meson excitations
$\phi^{'}$ \cite{san1}. The scalar field $\chi$ and
$F_{\phi}$ have been calculated in the mean field approximation 
and quark- gluon, gluon- gluon and quark- meson excitations are
treated perturbatively. 

With the above ansatz, the quark masses now become density dependent
through \( F_\phi \). It has been shown earlier \cite{san1} that for 
three flavour matter only \( <\vec\pi^{2}> \) develops non zero
value and \( <K^2> \) = \( <\eta^2> \)= 0 which means that strange quark mass
remains constant in the
medium. The u and d quark masses decrease in the medium with increase in
density. The chemical equilibrium and charge neutrality among the 
constituents implies \( \mu_{d(s)} = \mu_{u}+ \mu_{e} \) and 
\( (2/3) n_{u} - (1/3) n_{d} -(1/3)
n_{s} - n_{e}=0 \) respectively.  Baryon density
\( n_{B}=(1/3)\sum_{i}(n_{i}) \) where i= u,d,s. With these conditions we
calculate the thermodynamic potential up to second order in quark-gluon
interaction. The parameter set used in the present paper are 
\( B^{1/4} = 152 \) MeV, \( m_{u,d} = 92 \) MeV, \( m_s = 295 \) MeV, 
\( \alpha = 36\) MeV and strong coupling constant 
\( g_s (= 4 \pi \alpha_s) = 1.008 \).

\section{Rotating Star Solutions}

In this section we are going to have a brief discussion of the procedure 
of the rotating star calculations followed by the discussion of our 
results obtained. In figure 1 we have plotted the EOS for CCDM along
with the bag model results for interacting quark matter with the same 
bag pressure and strong coupling constant as given for CCDM.
The EOS for CCDM is found to be soft compared to that for the bag model. 

Once the EOS is obtained the next job is to solve the Einstein's 
equations for the rotating stars using the EOS. To solve the Einstein's 
equations we follow the procedure adopted by Komatsu {\it et.al.} \cite{a17}.
 In this 
work we briefly outline some of the steps only. The metric for a 
stationarily rotating star can be written as \cite{a18}
\be
ds^2 = -e^{\gamma+\rho} dt^2 + e^{2\alpha} \left(dr^2 + r^2 d\theta^2\right)
+e^{\gamma-\rho} r^2 sin^2\theta \left(d\phi - \omega dt\right)^2
\ee
where $\alpha$, $\gamma$, $\rho$ and $\omega$ are the gravitational potentials
which depend on $r$ and $\theta$ only. The Einstein's equations for 
the three potentials
$\gamma$, $\rho$ and $\omega$ have been solved by Komatsu {\it et.al.} using 
Green's function technique. The fourth potential $\alpha$ has been determined 
from other potentials. All the physical quantities may then be determined 
from these potentials \cite{a18}. 

Solution of the potentials, and hence the calculation of physical quantities, 
is numerically quite an involved process. There are several numerical codes 
in the community for this purpose. In the present paper, using the 
{\bf rns} code, developed by Stergioulas {\it et. al.}, we have studied 
the properties of rotating quark star. The results are discussed below. 

Let us first look at the constant $\Omega$ (angular velocity) sequences. 
In figure 2 we have plotted mass as a function of radius for different 
angular velocities starting from the static up to the keplerian limit. For 
comparison, the $M-R$ curves for both CCDM and bag model have been plotted 
in the same figure. From this figure one can see that, in the static limit, 
for the same mass the CCDM gives a much smaller radius compared to the bag 
model. For example, in the static case, for a star with mass 
$1.6 M_\odot$, the radius of a star is about 
11 Km for bag model compared to about 9.5 Km for the CCDM.
However, this difference is much more 
pronounced for rotating stars especially as we move towards the keplerian 
limit. For a star moving with keplerian frequency, in fact, there is no 
overlap between the mass-radius plots for the two models. As we vary the 
energy density from $8 \times 10^{14} gms/cm^3$ to $1.5 \times 10^{15} 
gms/cm^3$ the CCDM results in a mass range of $2.1 M_\odot - 2.5 M_\odot$ 
and the radius varies from 14 Km to 15 Km. For the same variation 
of energy density the bag model results in a mass variation from 
$2.9 M_\odot$ to $3.2 M_\odot$. The radius varies from 16.5 Km to 
17.8 Km for such stars. However these results are better explained in 
figure 3 as discussed below. 

In figure 3 we have plotted the mass as a function of central energy 
density. For $\epsilon_c = 1.2 \times 10^{15} gms/cm^3$ the mass of a 
static star is about $1.3 M_\odot$ in CCDM. For the same central 
energy density the bag model results in a star of mass $2 M_\odot$. 
So there is a huge difference between the two models even in the 
static case. Let us now look at the other extreme limit. When the star 
is rotating with keplerian frequency a central density of $1.2 \times 
10^{15} gms/cm^3$ results in a star of mass $3.1 M_\odot$ in the bag 
model and $2.2 M_\odot$ in the CCDM. 

The difference of the results obtained from the two models, as discussed  
above, can be ascribed  to the difference in the EOSs of the two models. 
The softer EOS of CCDM allows a star to have smaller 
size compared to that in the bag model for the same mass. 

Apart from the mass-radius relationship, one of the ways to look at the 
validity of an EOS is to look at the QPOs 
or the ISCO frequencies. For conventional rotating neutron stars the ISCO lies 
inside the star. However, this is not true for a rotating quark star. 
The ISCO frequencies are observable quantities and hence one can put 
a constraint on the EOS from these results. We have studied the 
ISCO frequencies for both the models for different rotational frequencies.
In figure 4 the QPO frequencies ($\Omega_+$) have been plotted as a 
function of mass of the star, rotating with keplerian angular velocity,
for the CCDM as well as the bag model.
From figure 4 we can see that for both the models the ISCO frequency increases 
with mass. The ISCO frequencies obtained from the 
CCDM are found to be somewhat higher compared to the bag model   
for the same range of $\epsilon_c$ as mentioned earlier. 
In the case of CCDM the range of ISCO frequency is $1263 Hz$ to $1490 Hz$ 
where as for the bag model it is $1101 Hz$ to $1351 Hz$. It should be noted
that for all the cases mentioned here, ISCO lies above the stellar surface.
Furthermore, especially for the CCDM, reducing the rotational frequency of the star 
results in the vanishing of the gap between the ISCO and the stellar surface.

Rossi X-ray Timing Explorer (RXTE) has observed kHz QPOs in around 20 LMXBs,
mostly showing double peaks. The high frequency peak, which corresponds to the
the ISCO frequency lies in the range 500 - 1200 Hz \cite{vander}. 
Our results are comparable to the higher end of observed ranges.
We would also like to mention here that the presence of crust for a star
in CCDM causes the gap to vanish even for keplerian rotational frequency.

In figure 5 we have plotted the ISCO frequency as a function of the 
rotational frequency. As in figure 4, the rotational frequency is the 
keplerian frequency. It can be seen that for both the models 
the ISCO frequency increases with rotational frequency. Furthermore, star in
bag model 
has lower $\Omega_+$ compared to that in CCDM for the same value of $\Omega$.

\section{Summary and Discussion}

We have studied the rotating compact stars within the 
framework of CCDM. This is the first time that CCDM has been employed 
towards the study of the rotating stars. The results have been compared 
with the bag model for same values of bag pressure and strong coupling.
We have found that the 
results obtained from the CCDM is very different compared to those 
obtained in the bag model. The size of a star obtained in the bag 
model is much more compared to that in the CCDM for the same 
mass. Furthermore, the difference gets much more pronounced as one moves 
towards the keplerian limit. We have also studied the ISCO frequencies 
for these two models. For both the models $\Omega_+$ increases with M. 
The $\Omega_+$ is higher in CCDM compared to that in the bag model. 
In our model, only the QPOs having frequency higher than 1260 Hz can be 
a strange star. 

There are several issues that remains to be addressed so far as the 
existence of the strange stars is concerned. Though it is true that
the quark models can be constrained using the observed ISCO frequencies,
while constructing the quark matter EOS, one should also
try to incorporate the essential QCD symmetries into the model. This
puts a further constraint on the quark matter EOS as it induces a qualitative 
difference in the EOS. For example, the bag model
EOS can be easily put into the form \( P= a (\rho - \rho_0) c^2/3 \)
\cite{steri2} where 
\( a \) is a positive number. In contrast, the best possible fit for the 
CCDM model EOS is given by the polynomial form \( P = a_0 (\rho - \rho_0) + 
a_1 (\rho - \rho_0)^{3/2} \). The coefficients are 
\( a_0 = 2.7983 \times 10^{20} \) and \( a_1 = 1.2245 \times 10^{12} \) 
with \( \rho_0 \approx 5.7 \times 10^{14} \)gms/cm\(^3 \). 

Moreover, bag model EOS being stiffer, which resulted in a higher mass for 
the static star, people were in search of a softer model to yield a lower value 
of the star mass. It now seems that a stiffer EOS is preferable to 
reproduce the observed QPO frequencies. One should make a further 
survey of the models at this stage and look for better EOSs which can 
reproduce both the star mass and the QPO frequency satisfactorily.

{\bf Acknowledgements}
AB would like to thank University Grants Commission for partial support 
through the grant PSW-083/03-04.

\newpage
\begin{figure}[t]
\vskip-2.1cm
\hskip-1.76cm
\centerline{\psfig{file=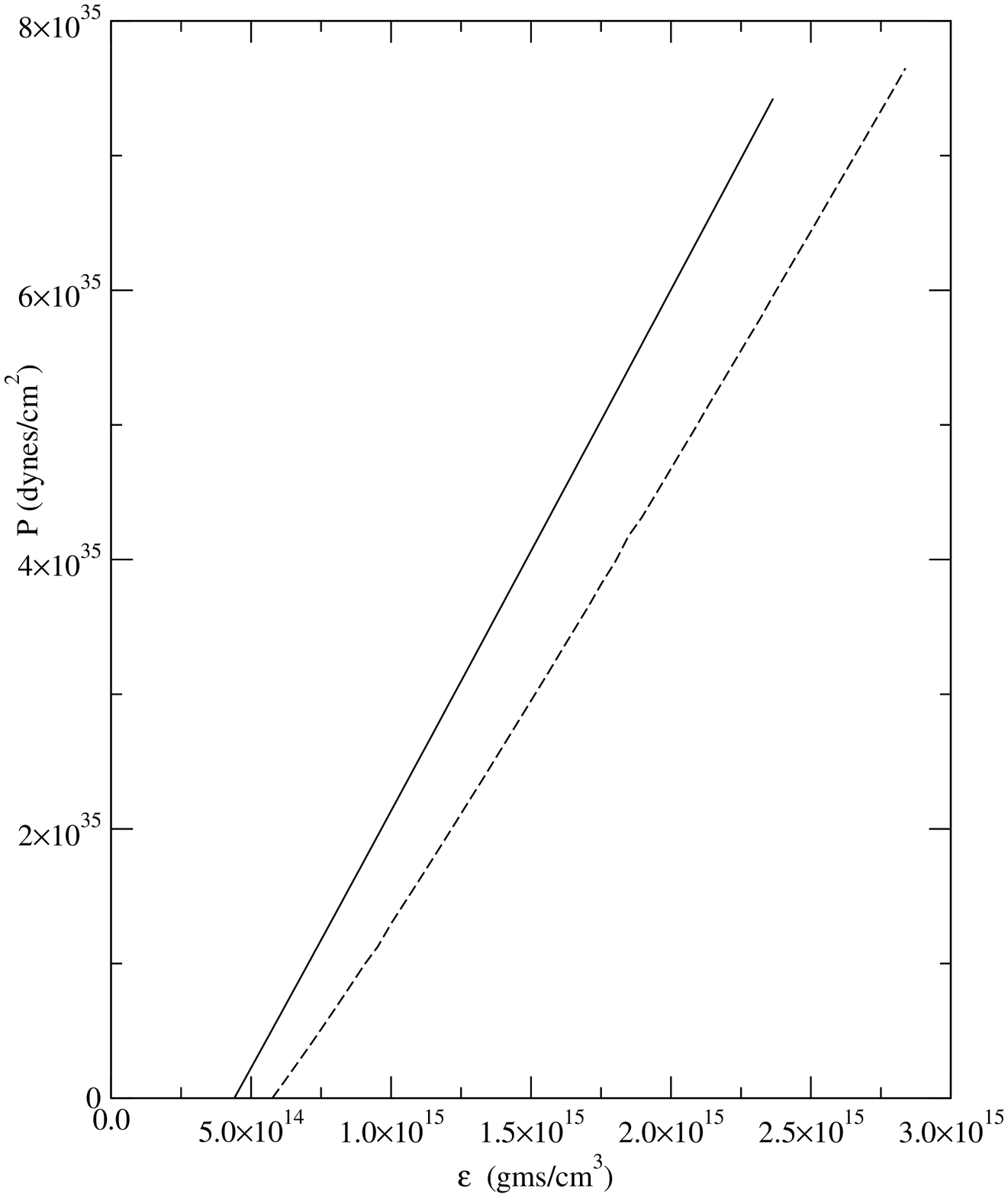,width=15cm}}
\caption{EOSs obtained from the Bag model (continuous line) and CCDM (dashed
line).}
\end{figure}
\newpage
\begin{figure}[t]
\vskip-2.1cm
\hskip-1.76cm
\centerline{\psfig{file=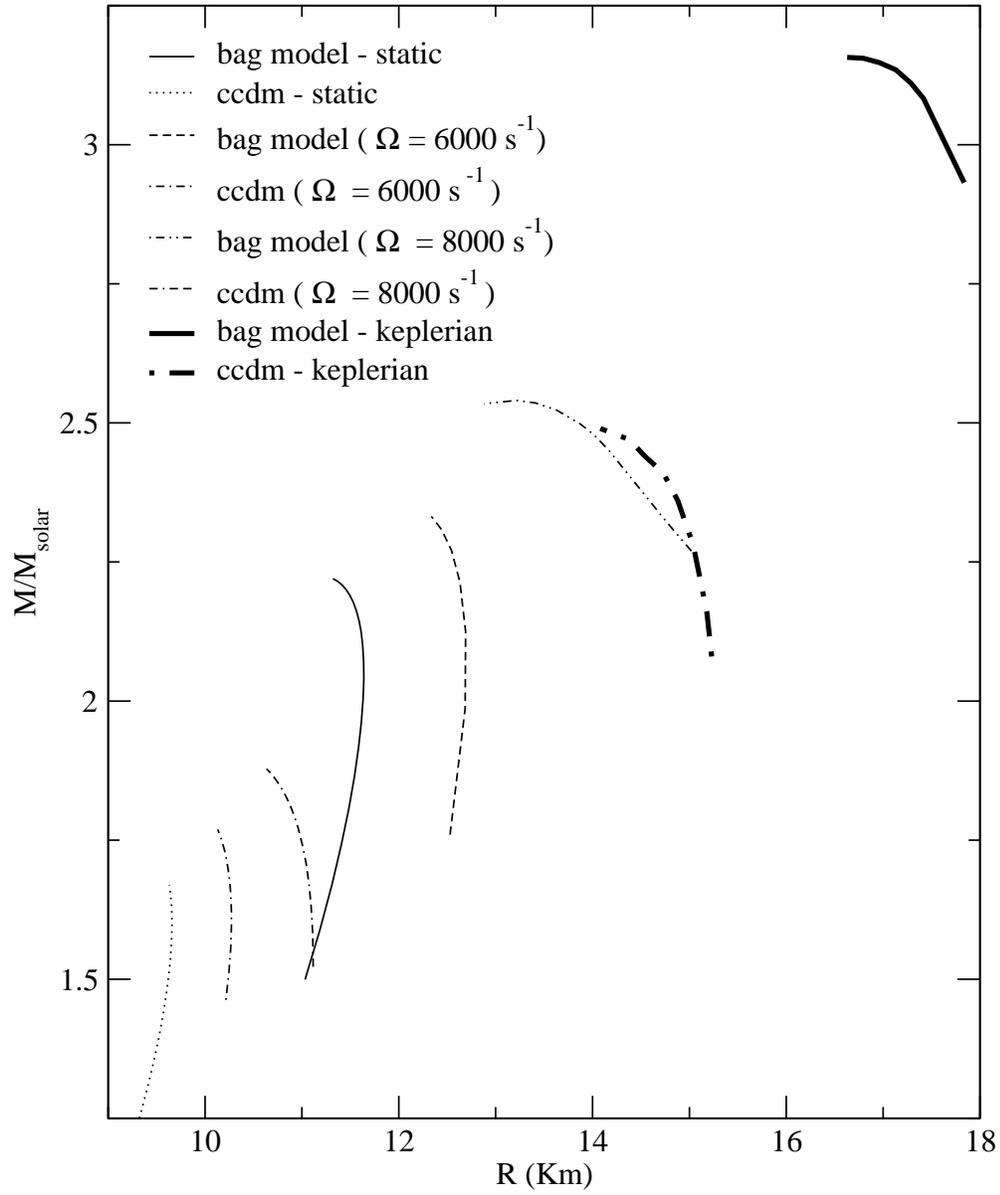,width=15cm}}
\caption{Mass-Radius plots at constant $\Omega$ for different values of
$\Omega$. }
\end{figure}
\newpage
\begin{figure}[t]
\vskip-2.1cm
\hskip-1.76cm
\centerline{\psfig{file=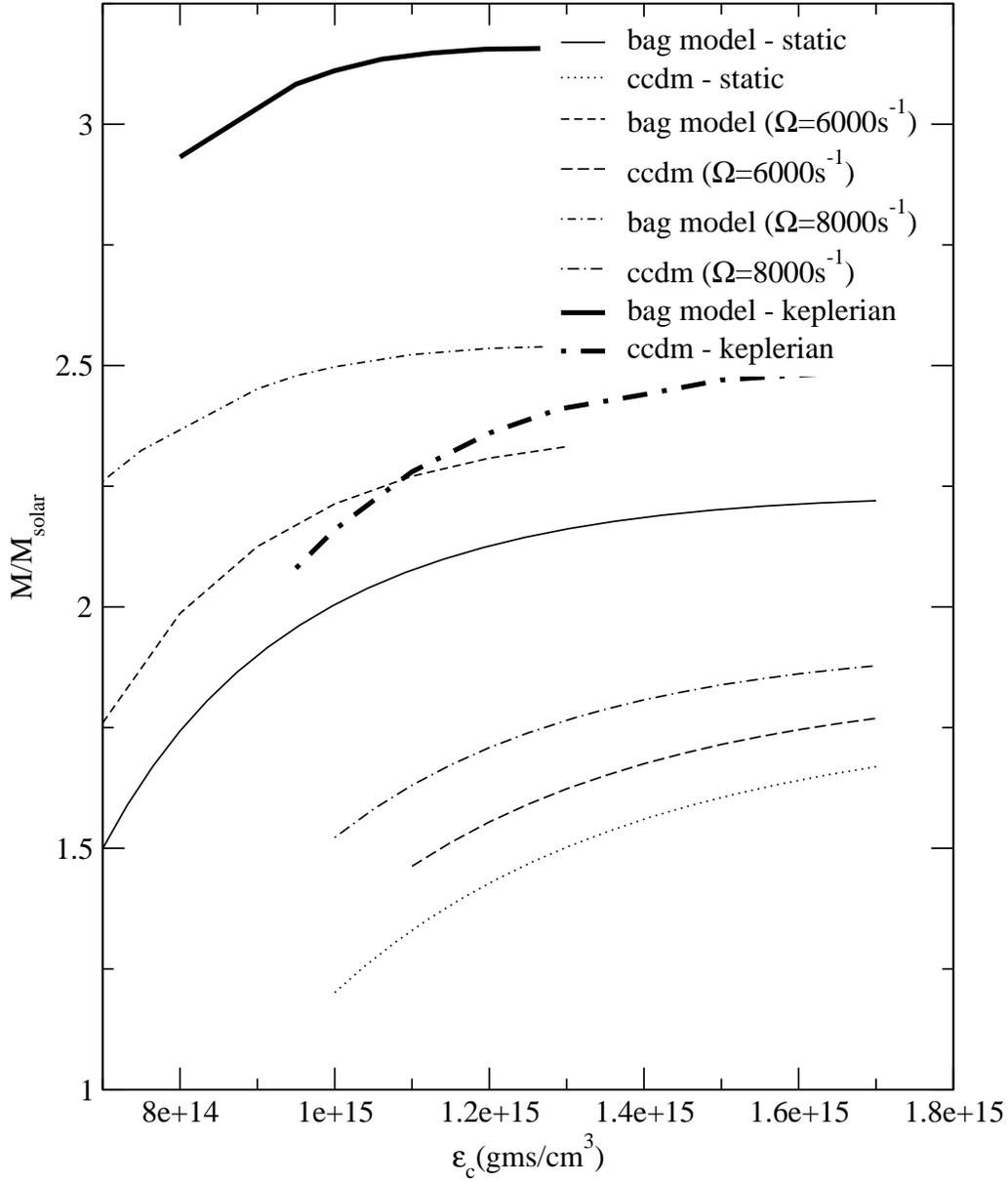,width=15cm}}
\caption{Mass-Central density plots for constant $\Omega$ for different
values of $\Omega$.}
\end{figure}
\newpage
\begin{figure}[t]
\vskip-2.1cm
\hskip-1.76cm
\centerline{\psfig{file=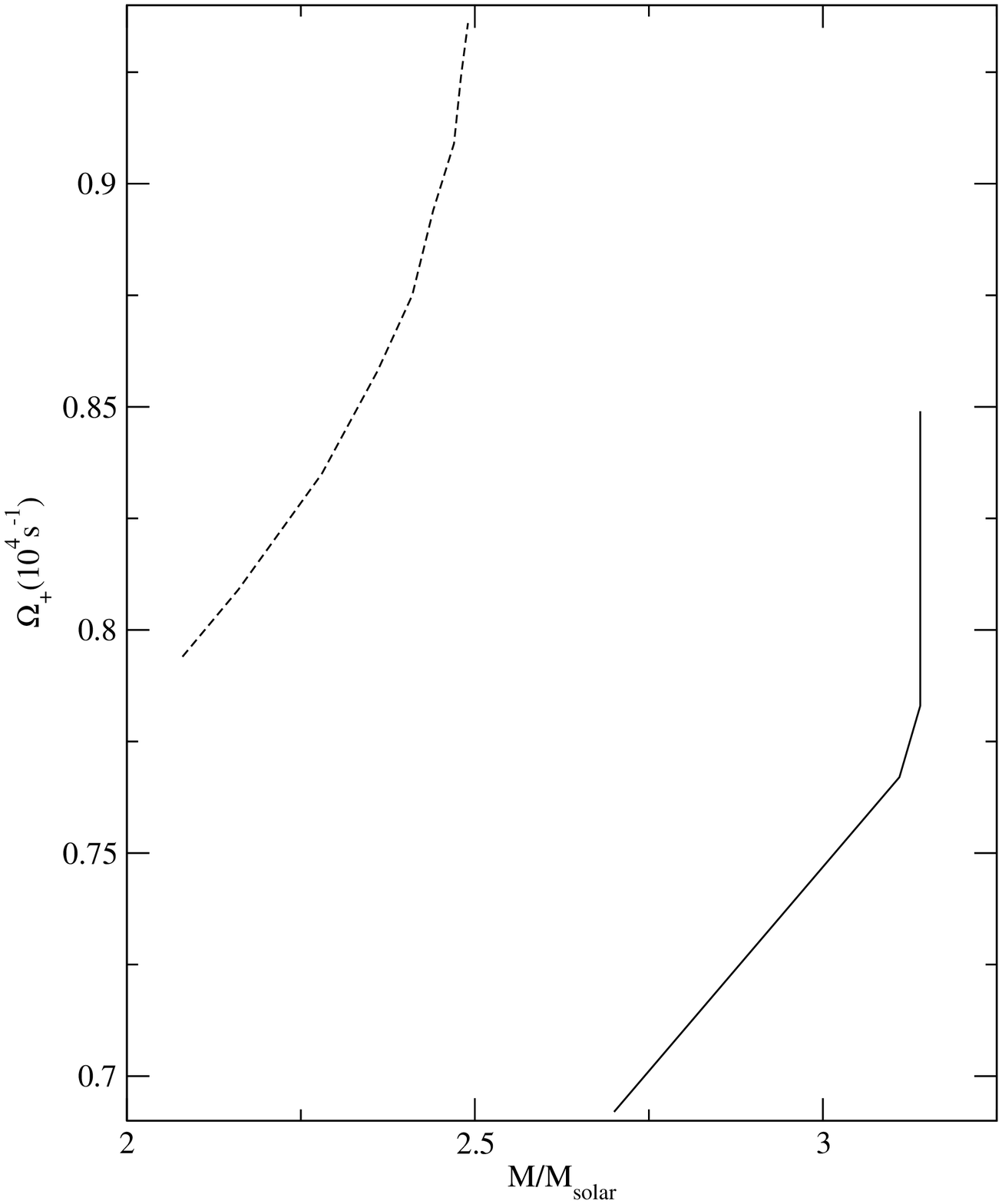,width=15cm}}
\caption{$\Omega_+$ as a function of mass for bag model (continuous line)
and for ccdm (dashed line).}
\end{figure}
\newpage
\begin{figure}[t]
\vskip-2.1cm
\hskip-1.76cm
\centerline{\psfig{file=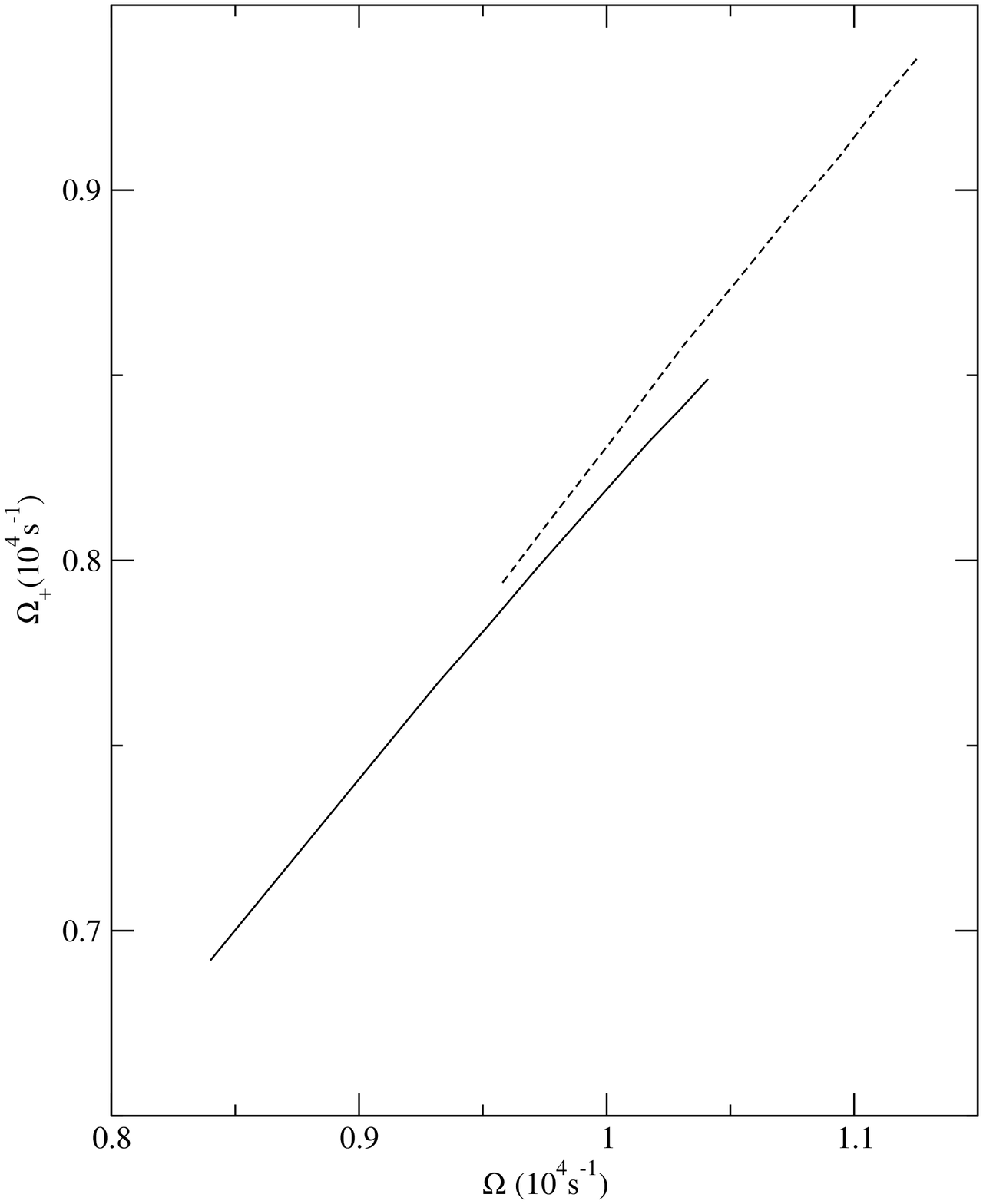,width=15cm}}
\caption{$\Omega_+$ as a function of $\Omega$ for bag model (continuous line)
and for ccdm (dashed line).}
\end{figure}
\end{document}